# Rock Classification in Petrographic Thin Section Images Based on Concatenated Convolutional Neural Networks

Cheng Su[1*], Sheng-jia Xu[1], Kong-yang Zhu[2], and Xiao-can Zhang[1]

[1]Institute for Geography & Spatial Information, School of Earth Sciences, Zhejiang University, Hangzhou, China.

[2]Institute of Geology, School of Earth Sciences, Zhejiang University, Hangzhou, China.

*Abstract*— **Rock classification plays an important role in rock mechanics, petrology, mining engineering, magmatic processes, and numerous other fields pertaining to geosciences. This study proposes a concatenated convolutional neural network (Con-CNN) method for classifying the geologic rock type based on petrographic thin sections. Herein, plane polarized light (PPL) and crossed polarized light (XPL) were used to acquire thin section images as the fundamental data. After conducting the necessary pre-processing analyses, the PPL and XPL images as well as their comprehensive image (CI) were incorporated in three convolutional neural networks (CNNs) comprising the same structure for achieving a preliminary classification; these images were developed by employing the fused principal component analysis (PCA). Subsequently, the results of the CNNs were concatenated by using the maximum likelihood detection to obtain a comprehensive classification result. Finally, a statistical revision was applied to fix the misclassification due to the proportion difference of minerals that were similar in appearance. In this study, 13 types of 92 rock samples, 196 petrographic thin sections, 588 images, and 63504 image patches were fabricated for the training and validation of the Con-CNN. The five-folds cross validation shows that the method proposed provides an overall accuracy of 89.97%, which facilitates the automation of rock classification in petrographic thin sections.**

## I. Introduction

Rock classification is essential for geological research, and plays an important role in numerous fields such as rock mechanics, petrology, mining engineering, magmatic processes, and applications associated with geosciences [1-3]. This classification can be accomplished via the characterization of different minerals in rocks, which is performed by using methods such as polarized light microscopy, X-Ray Diffraction (XRD), X-Ray Fluorescence (XRF), Atomic Absorption Spectroscopy (AAS), Electron Micro Probe Analyzer (EMPA), Scanning Electron Microscopy-Energy Dispersive X-ray spectroscopy (SEM-EDX), Transmission Electron Microscopy (TEM) [2]. Among these, the most important and widely used methodology is the manual analysis conducted by geologists on the image of the petrographic thin section, which is obtained by using the polarizing microscope. Due to the distinct optical properties of minerals, the thin section image can provide abundant petrographic information. With the rapid development in technologies associated with computer image processing in the past decades, a few attempts have been made to achieve an automatic elucidation of the rock and mineral information from the thin section image by using computer algorithms [4]. This approach demonstrates an enhanced efficiency, accuracy, and objectivity when compared to the traditional manual analysis. These analyses can be divided into three categories: (1) pore information extraction, (2) mineral identification, and (3) rock classification.

*A. Pore information extraction*
Pore is the void space between minerals in a rock, which is extremely critical for petroleum and gas exploration. It is common practice to extract pore information such as the geometric shape, size, type and co-ordination number. These parameters identify and measure the pore spot in cast thin-section images, which are thin-sections impregnated with colored epoxy. Based on their unique color, pores can be identified by threshold methods in the RGB or HSV color spaces [5, 6]. In addition, pattern recognition and GIS-based methods are applied to extract the boundary and region of the pore as a polygon object, and, further, to quantitatively calculate its shape, orientation, type, and spatial distribution [7-10]. Moreover, the deep-learning methods classify the thin-section image pixel by pixel, which is called image semantic segmentation, creating the labeled output image, where every single labeled pixel represents a mineral class or pore [11, 12]. After that, the extracted pore information can be used to estimate rock permeability and anisotropy in reservoir simulation, hydrology, and environmental engineering [13-15].

*B. Mineral identification*
Similar to pore information extraction, the basic theory of mineral identification is that different minerals exhibit specific colors and textures due to their optical properties. [16] summarized various image processing and pattern recognition techniques devoted to this field. Some other works use machine learning methods, such as artificial neural networks (ANN), support vector machine (SVM), U-Net (a kind of convolutional neural network, CNN) and instance segmentation, to perform intelligent mineral identification using computer statistical analysis under human expert supervision [1, 11, 17-20]. After separating diverse mineral types, some other works focus on extracting mineral grain geometric parameters, like boundary, shape, size and percentage [21-24].

*C. Rock classification*
To classify the rocks into geologic types, it is necessary to develop effective methods for characterizing the thin-section image features. [25] used 23 texture features as an input to construct an artificial neural network (ANN) for classifying carbonate rocks into mud-stone, wackestone, packstone and grainstone. [26] used 13 color features to classify nine kinds of rocks. They compared different color-spaces and pattern recognition methods and discovered that the best options were utilizing the methodologies of CIELab and nearest neighbor (NN). [27] calculated color histogram and edge and used an unsupervised cluster method to classify the intrusive igneous rock thin section images into diorite, tonalite, granite, and adamellite. [3] proposed a transfer learning method for sandstone classification. The advantage of their method is that the well-trained model can be applied to any other untrained domain with little manual labeling effort, irrespective of the huge differences among sandstones in different domains. [28] designed an image database of microscopic rock for searching visually similar images. It can be considered as rock classification by cluster method. With a similar purpose, some works classify rocks by image features rather than geologic types [29-31].

Currently, given the support of efficient computing equipment and big data, deep learning methods are applied in many fields. They have the potential to extract features and relations through training and learning from a large sample, and to provide a data-driven solution without human engineers. Among them, convolutional neural networks (CNN) are widely used in image classification, image understanding, and many other image processing fields, because their computing can maintain the local topological properties of images [32]. In this study, a convolutional neural network with concatenated structure is proposed for general geologic type rock classification through the petrographic thin section image. A total of 6000 images acquired from 106 rock samples of 13 rock types have been used for training and testing. The 5-fold cross-validation result shows the overall accuracy is 89.97% and Kappa coefficient is 0.86

## II. MATERIALS

This study covers a total of 13 types (Andesite, Granite, Peridotite Gabbro Rhyolite Tuff Diorite Phonolite Basalt Syenite Limestone Sandstone Schist) of 92 representative rocks in igneous, sedimentary and metamorphic categories. The rock samples are mainly from the GeosecSlides and the Classic North American Rock Collection of ward's science. The remaining part of samples are provided by School of Earth Sciences, Zhejiang University. 296 thin sections made of these rocks were collected and 2208 of their digital images were taken under the AxioCamMR5 microscope. All images were captured in RGB format at 2.5x optical magnification with a dimension of 2688×2016 pixels with 150 dpi resolution. Exposure parameter and white balance were automatically adjusted during the acquisition so that all image results can be displayed as close as possible to the observations under the microscope. We used both plane polarized light (PPL) and crossed polarized light (XPL) to take the photos to obtain more comprehensive feature information. Considering that different minerals have different extinction properties, we also captured these thin section images in multiple rotation angles.

## III. METHOD

For any image, there are only color and texture information represented via pixel value and spatial arrangement. This is the fundamental of object identification, classification, and image understanding, performed by computer or human beings. However, two special phenomena should be noted in petrographic thin section images. One is the complex combination of the optical properties of minerals: the PPL images show the shape, color and cleavage while the XPL images show the extinction and interference of minerals (see **Figure 1**). The second phenomenon is the distribution of minerals: some rocks are similar in composition and appearance but present different proportions of minerals. For instance, both granite and diorite contain quartz, feldspar and dark minerals such as biotite, amphibole and magnetite, but the former has more quartz while the latter presents more dark minerals (see **Figure 2**).

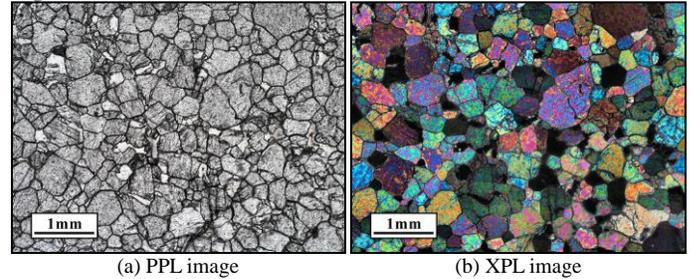

(a) PPL image      (b) XPL image
**Figure 1 A sample of petrographic thin section image (peridotite).**

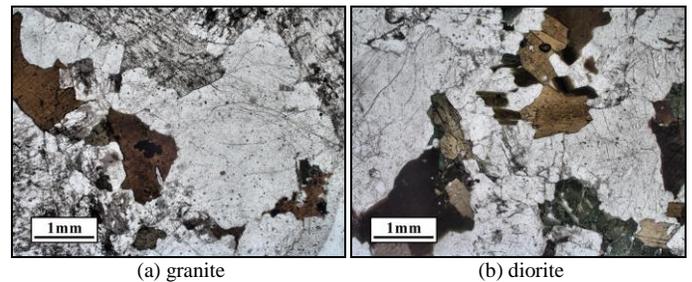

(a) granite      (b) diorite
**Figure 2 Granite and diorite petrographic thin section images.**

According to the characteristics of petrographic thin section images, a concatenated convolutional neural network (Con-CNN) was proposed in this paper (see **Figure 3**). The core technology of this method is CNN for extracting particular features of images. After using a large number of samples for learning and training, the CNN could establish an appropriate mapping relationship between the input image and the output category, which is the classification model we wanted. There are three independent CNN branches in the Con-CNN to classify the input image into the corresponding geologic rock type, taking full advantage of PPL, XPL and their synthetic image of one petrographic thin section, respectively. The results of these three branches are concatenated by using maximum likelihood to get a comprehensive classification result. In the end, in order to obtain the final output, a statistical revision is performed to fix misclassifications due to the proportion difference of minerals that are similar in appearance. The whole Con-CNN method can be divided into the five parts detailed below.

### Part I. Image Pre-process

Before training and classification, some pre-processing steps need to be applied on the input raw data for the obtention of a more stable performance. First, the PPL and XPL original images were enhanced by the histogram equalization method [33], increasing the intensity range by redistributing the pixel levels. This can enhance the contrast of images and avoid pixels clustered around the middle of the available range of intensities, yielding detailed information at the same brightness level. Second, a composite image with six layers was created via layer stacking PPL and XPL (three layers: $PPL_R$, $PPL_G$, $PPL_B$, and same for XPL). Then the composite image was transformed by the Principal Component Analysis (PCA) [33], which is a statistical procedure using an orthogonal transformation to convert the raw data into a set of values of linearly uncorrelated principal components (PC). The PCA transformation is such that information moves rapidly forward to the PCs in front. Therefore, the first three PC, representing most of the original information, were selected as RGB layers for the construction of a comprehensive image (CI), which is a fusion of the PPL and XPL. In the end of the pre-processing steps, three images (PPL, XPL and CI) needed to be normalized to keep pixel values at the same dimension of quantity.

*Part II. Image Slicing*

In an image, pixel value and spatial arrangement combine to produce information. In other words, image features are represented by some adjacent local pixels, meaning that they are scale dependent. On a large scale, zooming in on parts of the image reveals more local details and there is more discrimination between images. On the contrary, zooming out to view the whole image, there are more general characteristics but fewer details. Some petrographic thin section images are very visually similar. Therefore, the classification method should take full advantage of image details. In this paper, the original image acquired from the microscope presents 2688×2016 pixels, and was regularly sliced into 12×9 patches with 224×224 pixels. Each patch was taken as an independent image for training and classification. Although the small patch image can better represent local details, the overall features of the original whole thin section image are fragmented and lost. This issue will be considered in Part V below.

*Part III. Convolutional Neural Network*

When the input data were prepared, a CNN was constructed as three independent modals for classification of each type of input image (as shown in **Figure 3Figure 3**). The basic structure of the CNN part proposed in this paper refers to the classical model of LeNet [34] and VGG16 [35]. The input image is passed through five feature extraction blocks (FEB) and three mapping blocks (MB) to be classified into one of 13 rock types. The FEB contains three computations: the first is a convolutional layer, where filters with small receptive fields of 3×3 are applied with a stride of 1 pixel and none of padding. The second is a pooling layer, where a 2×2 Max-pooling operation is performed with a stride of 2 pixels following the convolutional layer. The third is an activation function, where the pooling layer result is equipped with the rectification non-linearity (ReLU). After 5 FEBs, the size of the input image gradually changed from 224×224 to 112×112, 56×56, 28×28, 14×14, and finally to 7×7. The MB contains two computations: the first is fully connected layers, where all nodes between layers are connected, similar to a classic artificial neural network (ANN), to establish mapping relations. The second is an activation function using the sigmoid function. It is noted that the last MB is a SoftMax layer, which could map multiple values to the probability of belonging to a category.

*Part IV. Concatenation*

For one petrographic thin section, three images (PPL, XPL and CI) have been used as input data for the CNN for classification. Each classification result is the probability of belonging to one of the 13 rock types. Here, the concatenation was applied to integrate these three CNN classification results. More specifically, these three results were weighted and averaged to get the final result. Through many experiments, we suggest that weights are 0.4 for PPL, 0.4 for XPL, and 0.2 for CI. For instance, if granite probability from the PPL image is 0.92, from the XPL is 0.87, and from the CI is 0.89, then the final probability of belonging to granite is 0.94×0.4+0.87×0.4+0.89×0.2=0.902. After the concatenation, the final rock type is set as the one with the highest probability, using the maximum likelihood principle.

*Part V. Statistical Revision*

As described in Part II, the original image was sliced into small image patches for classification. For one petrographic thin section, 324 image patches (the PPL image was sliced into 12×9=108 patches, and the same was done for XPL and CI) have been classified. After each patch has been classified, the mode of rock types was calculated to find out which rock type is more frequent among the 324 classification results. Then the most common rock type was uniformly assigned to all small image patches belonging to the same petrographic thin section. This revision restores the overall statistical characteristics of the original image to some extent and avoids the issue of original information fragment and loss caused by image slicing.

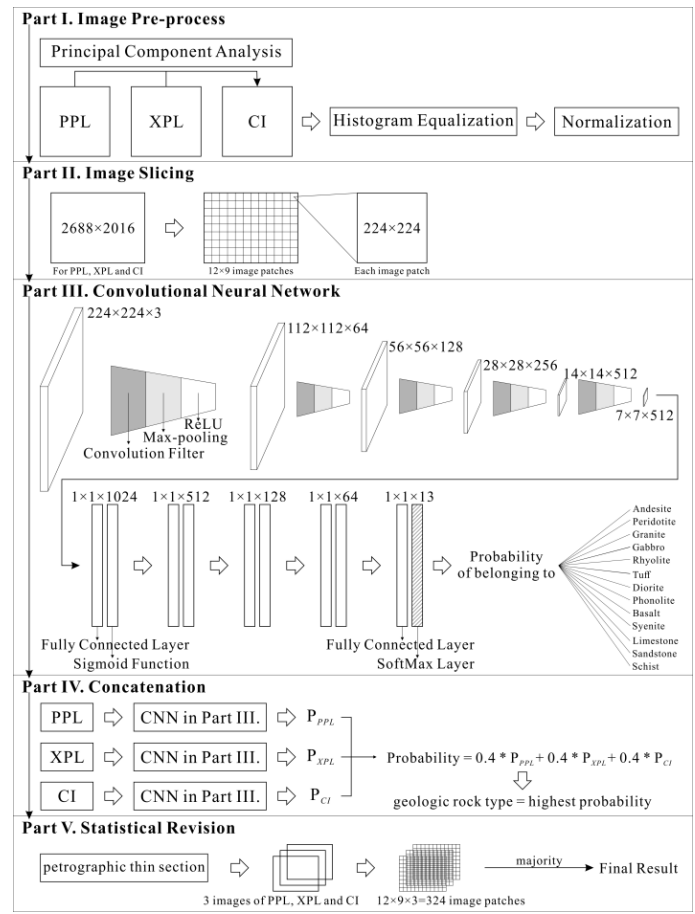

**Figure 3 Architecture of concatenated convolutional neural network (Con-CNN).**

## IV. EXPERIMENTS AND DISCUSSION

To evaluate the proposed Con-CNN model, a test experiment was implemented by programming with PyTorch deep learning framework [36]. Petrographic thin sections for a total of 92 rock samples of 13 types were prepared. A total of 2208 images, acquired with a microscope for accurate rock type information, were sliced into 238464 small image patches. All the image patches were used as training and validation samples and put into the Con-CNN for modeling. To get the general performance of the Con-CNN model, a 5-fold cross validation was applied by randomly dividing the all image samples into four training groups and one validation group. This experiment was repeated five times to get the average evaluation result. For each experiment in cross validation, after about 20,000 iterations the model tends to converge. **Table 1** shows the average classification performance of the Con-CNN model, which presents an overall accuracy of 89.97% and a Kappa coefficient of 0.86.

**Table 1 Confusion matrix.**

|  | andesite | basalt | diorite | gabbro | granite | limestone | peridotite | phonolite | rhyolite | sandstone | schist | syenite | tuff |
|---|---|---|---|---|---|---|---|---|---|---|---|---|---|
| andesite | 1.00 | 0.00 | 0.00 | 0.00 | 0.00 | 0.00 | 0.00 | 0.00 | 0.00 | 0.00 | 0.00 | 0.00 | 0.00 |
| basalt | 0.00 | 0.98 | 0.00 | 0.00 | 0.00 | 0.00 | 0.02 | 0.00 | 0.00 | 0.00 | 0.00 | 0.00 | 0.00 |
| diorite | 0.00 | 0.00 | 0.58 | 0.00 | 0.42 | 0.00 | 0.00 | 0.00 | 0.00 | 0.00 | 0.00 | 0.00 | 0.00 |
| gabbro | 0.00 | 0.00 | 0.02 | 0.75 | 0.00 | 0.16 | 0.00 | 0.04 | 0.00 | 0.00 | 0.00 | 0.02 | 0.00 |
| granite | 0.00 | 0.00 | 0.00 | 0.00 | 0.61 | 0.14 | 0.00 | 0.00 | 0.00 | 0.00 | 0.00 | 0.11 | 0.14 |
| limestone | 0.00 | 0.00 | 0.00 | 0.00 | 0.00 | 0.98 | 0.00 | 0.01 | 0.00 | 0.00 | 0.00 | 0.00 | 0.00 |
| peridotite | 0.00 | 0.00 | 0.00 | 0.00 | 0.00 | 0.00 | 0.99 | 0.00 | 0.00 | 0.00 | 0.00 | 0.00 | 0.00 |
| phonolite | 0.00 | 0.00 | 0.00 | 0.00 | 0.00 | 0.00 | 0.00 | 1.00 | 0.00 | 0.00 | 0.00 | 0.00 | 0.00 |
| rhyolite | 0.00 | 0.00 | 0.00 | 0.00 | 0.00 | 0.00 | 0.00 | 0.00 | 0.98 | 0.00 | 0.00 | 0.00 | 0.01 |
| sandstone | 0.00 | 0.00 | 0.00 | 0.00 | 0.03 | 0.00 | 0.00 | 0.02 | 0.00 | 0.63 | 0.00 | 0.31 | 0.00 |
| schist | 0.00 | 0.00 | 0.00 | 0.00 | 0.00 | 0.00 | 0.00 | 0.00 | 0.00 | 0.00 | 1.00 | 0.00 | 0.00 |
| syenite | 0.00 | 0.00 | 0.02 | 0.12 | 0.00 | 0.01 | 0.00 | 0.02 | 0.00 | 0.00 | 0.00 | 0.82 | 0.00 |
| tuff | 0.00 | 0.00 | 0.00 | 0.00 | 0.00 | 0.00 | 0.00 | 0.00 | 0.00 | 0.00 | 0.00 | 0.00 | 1.00 |

Three comparison experiments were implemented: the first one only uses the PPL image for training and classification. The second one augments the image input data by random rotation, translation, and scaling, which is a widely used method to expand training samples in machine learning for imaging studies. The third one replaces the CNN part of the architecture proposed in this paper by a deeper CNN of ResNet-50 [37]. These three experiments were all done through a 5-fold cross validation with the same materials. **Figure 4** shows their ROC and AUC results. The comparison results show that none of the three comparison experiments work as well as the original method detailed in section 3. This may have been caused by the PPL image not being able to fully reveal the optical characteristics of minerals in petrographic thin sections, thus affecting the rock classification. A possible explanation for the results of the second test is that the data augmentation did not conspicuously improve the classification result. We believe that this is due to the fact that the petrographic thin sections color and texture are relatively uniform, random and stable, meaning that some samples could express their complete features, unlike the general photo with a special direction. Therefore, using random rotation, translation, and scaling processes to expand the training group of samples could not provide new information. For the third test, a deeper CNN did not significantly improve the classification accuracy, suggesting that the mapping relationship between petrographic thin section images and geologic rock types might be relatively simple and could be constructed through a CNN with relatively few layers.

## V. CONCLUSION

In this study, a Con-CNN method was proposed for rock classification based on petrographic thin section images. Two images were acquired with plane polarized light and crossed polarized light, respectively, and their comprehensive image was fused by the principal component analysis method. Subsequently, this image was divided into small patches and incorporated in the CNN for classification. After the concatenation of the CNN results and the statistical revision process had been completed, the suggestion of the final rock type was successfully obtained. The experiments showed that the Con-CNN method could effectively extract petrographic thin section image features and establish the mapping relationship between the image of a particular rock type and its geologic characteristics via learning from samples; further, this method exhibited a good performance. As a result, the Con-CNN provides an automated solution for rock classification in petrographic thin section images.

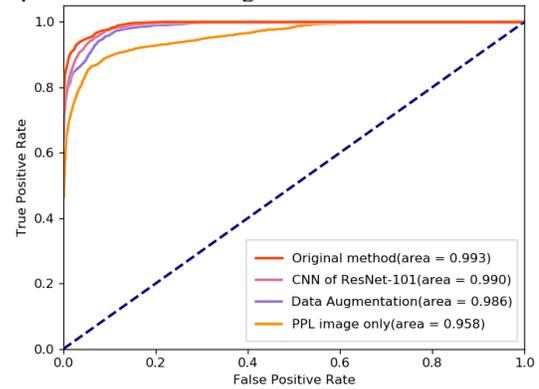

**Figure 4 ROC and AUC comparison.**

## ACKNOWLEDGMENT

This research is supported by the National Key R&D Program of China (NO. 2018YFB0505002), COMRA Major Project (NO. DY135-S1-01-01-03), and Zhejiang Provincial Natural Science Foundation of China (LY17D010006). Comments from anonymous reviewers are appreciated.